\documentclass[twocolumn,showpacs,preprintnumbers,amsmath,amssymb,prl,superscriptaddress]{revtex4}
\usepackage{graphicx}
\usepackage{dcolumn}
\usepackage{bm}

\begin{document}
\title{Three-particle coincidence of the long range pseudorapidity correlation in high energy nucleus-nucleus collisions}
\medskip
\affiliation{Argonne National Laboratory, Argonne, Illinois 60439, USA}
\affiliation{University of Birmingham, Birmingham, United Kingdom}
\affiliation{Brookhaven National Laboratory, Upton, New York 11973, USA}
\affiliation{University of California, Berkeley, California 94720, USA}
\affiliation{University of California, Davis, California 95616, USA}
\affiliation{University of California, Los Angeles, California 90095, USA}
\affiliation{Universidade Estadual de Campinas, Sao Paulo, Brazil}
\affiliation{University of Illinois at Chicago, Chicago, Illinois 60607, USA}
\affiliation{Creighton University, Omaha, Nebraska 68178, USA}
\affiliation{Czech Technical University in Prague, FNSPE, Prague, 115 19, Czech Republic}
\affiliation{Nuclear Physics Institute AS CR, 250 68 \v{R}e\v{z}/Prague, Czech Republic}
\affiliation{University of Frankfurt, Frankfurt, Germany}
\affiliation{Institute of Physics, Bhubaneswar 751005, India}
\affiliation{Indian Institute of Technology, Mumbai, India}
\affiliation{Indiana University, Bloomington, Indiana 47408, USA}
\affiliation{University of Jammu, Jammu 180001, India}
\affiliation{Joint Institute for Nuclear Research, Dubna, 141 980, Russia}
\affiliation{Kent State University, Kent, Ohio 44242, USA}
\affiliation{University of Kentucky, Lexington, Kentucky, 40506-0055, USA}
\affiliation{Institute of Modern Physics, Lanzhou, China}
\affiliation{Lawrence Berkeley National Laboratory, Berkeley, California 94720, USA}
\affiliation{Massachusetts Institute of Technology, Cambridge, MA 02139-4307, USA}
\affiliation{Max-Planck-Institut f\"ur Physik, Munich, Germany}
\affiliation{Michigan State University, East Lansing, Michigan 48824, USA}
\affiliation{Moscow Engineering Physics Institute, Moscow Russia}
\affiliation{City College of New York, New York City, New York 10031, USA}
\affiliation{NIKHEF and Utrecht University, Amsterdam, The Netherlands}
\affiliation{Ohio State University, Columbus, Ohio 43210, USA}
\affiliation{Old Dominion University, Norfolk, VA, 23529, USA}
\affiliation{Panjab University, Chandigarh 160014, India}
\affiliation{Pennsylvania State University, University Park, Pennsylvania 16802, USA}
\affiliation{Institute of High Energy Physics, Protvino, Russia}
\affiliation{Purdue University, West Lafayette, Indiana 47907, USA}
\affiliation{Pusan National University, Pusan, Republic of Korea}
\affiliation{University of Rajasthan, Jaipur 302004, India}
\affiliation{Rice University, Houston, Texas 77251, USA}
\affiliation{Universidade de Sao Paulo, Sao Paulo, Brazil}
\affiliation{University of Science \& Technology of China, Hefei 230026, China}
\affiliation{Shandong University, Jinan, Shandong 250100, China}
\affiliation{Shanghai Institute of Applied Physics, Shanghai 201800, China}
\affiliation{SUBATECH, Nantes, France}
\affiliation{Texas A\&M University, College Station, Texas 77843, USA}
\affiliation{University of Texas, Austin, Texas 78712, USA}
\affiliation{Tsinghua University, Beijing 100084, China}
\affiliation{United States Naval Academy, Annapolis, MD 21402, USA}
\affiliation{Valparaiso University, Valparaiso, Indiana 46383, USA}
\affiliation{Variable Energy Cyclotron Centre, Kolkata 700064, India}
\affiliation{Warsaw University of Technology, Warsaw, Poland}
\affiliation{University of Washington, Seattle, Washington 98195, USA}
\affiliation{Wayne State University, Detroit, Michigan 48201, USA}
\affiliation{Institute of Particle Physics, CCNU (HZNU), Wuhan 430079, China}
\affiliation{Yale University, New Haven, Connecticut 06520, USA}
\affiliation{University of Zagreb, Zagreb, HR-10002, Croatia}

\author{B.~I.~Abelev}\affiliation{University of Illinois at Chicago, Chicago, Illinois 60607, USA}
\author{M.~M.~Aggarwal}\affiliation{Panjab University, Chandigarh 160014, India}
\author{Z.~Ahammed}\affiliation{Variable Energy Cyclotron Centre, Kolkata 700064, India}
\author{A.~V.~Alakhverdyants}\affiliation{Joint Institute for Nuclear Research, Dubna, 141 980, Russia}
\author{B.~D.~Anderson}\affiliation{Kent State University, Kent, Ohio 44242, USA}
\author{D.~Arkhipkin}\affiliation{Brookhaven National Laboratory, Upton, New York 11973, USA}
\author{G.~S.~Averichev}\affiliation{Joint Institute for Nuclear Research, Dubna, 141 980, Russia}
\author{J.~Balewski}\affiliation{Massachusetts Institute of Technology, Cambridge, MA 02139-4307, USA}
\author{O.~Barannikova}\affiliation{University of Illinois at Chicago, Chicago, Illinois 60607, USA}
\author{L.~S.~Barnby}\affiliation{University of Birmingham, Birmingham, United Kingdom}
% ...
\collaboration{STAR Collaboration}\noaffiliation
 
\date{\today}
\begin{abstract}
We report the first three-particle coincidence measurement in pseudorapidity ($\Delta\eta$) between a high
transverse momentum ($p_{\perp}$) trigger particle and two lower $p_{\perp}$ associated particles within azimuth
$\mid$$\Delta\phi$$\mid$$<$0.7 in $\sqrt{{\it s}_{NN}}$ = 200 GeV $d$+Au and Au+Au collisions. Charge ordering properties are
exploited to separate the jet-like component and the ridge (long-range $\Delta\eta$ correlation).
The results indicate that the particles from the ridge are uncorrelated in $\Delta\eta$ not only with the trigger particle but also
between themselves event-by-event. In addition, the production of the ridge appears to be uncorrelated to the presence of the narrow jet-like component.
\end{abstract}
\maketitle
Di-hadron coincidence measurements provide a powerful tool to study the properties of the medium created in ultra-relativistic heavy-ion collisions. 
The observation of the long range pseudorapidity correlation in central Au+Au collisions~\cite{PRL}, called the ridge~\cite{joern}, where hadrons are correlated 
with a high transverse momentum ($p_{\perp}$) trigger particle in 
azimuth ($\Delta\phi$$\sim$0) but extended to large relative pseudorapidity ($\Delta\eta$), has generated great interest. 
Various theoretical models are proposed to explain this phenomenon, including 
(i) longitudinal flow push~\cite{nestor}, (ii) broadening of quenched jets in turbulent color fields~\cite{majumdar}, 
(iii) recombination between thermal and shower partons~\cite{hwa}, (iv) elastic collisions between 
hard and medium partons (momentum kick)~\cite{wong}, and (v) particle excess due to QCD bremsstrahlung or 
color flux tube fluctuations focused by transverse radial flow~\cite{voloshin,shuryak,dumitru,dusling,jun}. 
Models (i)-(iv) attribute the ridge to jet-medium interactions: particles from jet 
fragmentation in vacuum result in a peak at $\Delta\eta$$\sim$0 
and those affected by the medium are diffused broadly in $\Delta\eta$ forming the ridge.
Model (v) attributes the ridge to the medium itself, and its correlation with
high $p_{\perp}$ particles is due to the transverse radial flow.\\ 
\indent Despite very different physics mechanisms, all models~\cite{nestor,majumdar,hwa,wong,voloshin,shuryak,dumitru,dusling,jun} give 
qualitatively similar distributions of correlated hadrons with a high-$p_{\perp}$ trigger particle. Some of these model ambiguity can be
lifted by 3-particle coincidence measurements. We analyze the hadron pair densities from 3-particle coincidence measurements in 
($\Delta\eta_{1}$,$\Delta\eta_{2}$), the pseudorapidity differences between two 
associated particles and a trigger particle. We exploit charge combinations 
in an attempt to separate the jet-like and ridge components and study their distributions, without assuming the $\Delta\eta$ shape of the ridge. 
Jet fragmentation in vacuum should give a peak at ($\Delta\eta_{1}$,$\Delta\eta_{2}$)$\sim$(0,0), while
particles from the ridge would produce structures that 
depend on its physics mechanism. Correlation between particles from jet fragmentation and the ridge 
would generate horizontal or vertical stripes ($\Delta\eta_{1}$$\sim$0 or $\Delta\eta_{2}$$\sim$0) in the 3-particle coincidence measurement.\\
\indent Results are reported for minimum bias $d$+Au, peripheral 40-80$\%$ 
and central 0-12$\%$ Au+Au collisions at $\sqrt{{\it s}_{NN}}$ = 200 GeV from the STAR experiment~\cite{STAR}. 
The 40-80$\%$ data are from the minimum bias sample, and the 0-12$\%$ data are triggered by the 
Zero Degree Calorimeters (ZDC) in combination with the Central Trigger Barrel (CTB). 
This analysis uses 6.5 $\times$ 10$^{6}$ $d$+Au events taken in 2003, and 
6.0 $\times$ 10$^{6}$ peripheral and 1.9 $\times$ 10$^{7}$ central Au+Au events taken in 2004. 
The data are analyzed in finer centrality bins for Au+Au collisions~\cite{jason} and are combined for better statistics.\\
\indent The reconstructed event vertex is restricted within $\mid$$z_{\rm{vtx}}$$\mid<$30 $\rm cm$ 
along the beam line from the center of the STAR Time Projection Chamber (TPC)~\cite{tpc}, which sits in a uniform 
0.5 T magnetic field. The data were taken with both magnetic field polarities.
The trigger and associated particles are restricted to $\mid$$\eta$$\mid$$<$1 
and their  $p_{\perp}$ ranges are 3$<$$p_{\perp}^{\rm{(t)}}$$<$10 GeV/$c$ and 
1$<$$p_{\perp}^{\rm{(a)}}$$<$3 GeV/$c$, respectively. The correlated single and pair densities with trigger particle 
are corrected for the centrality-, $p_{\perp}$-, $\phi$-dependent 
reconstruction efficiency for associated particles and the $\phi$-dependent efficiency for 
trigger particles, and are normalized per corrected trigger particle.\\
\indent Due to the high TPC occupancy of Au+Au events, track pairs close in $\eta$ and
$\phi$ can be merged and reconstructed as single tracks. This results in
deficits in pair density at $\Delta\eta$$\sim$0 and at small, but non-zero, $\Delta\phi$ whose value depends on $p_{\perp}$, charge combination and magnetic field polarity. 
To reduce this effect, we apply cuts to exclude close track pairs in real and mixed events. 
Losses due to those cuts are compensated for by the acceptance correction obtained from mixed events.
To ensure the mixed events have similar characteristics as the real events, we mix events from the same centrality bin without requiring a 
trigger particle and with the same magnetic field polarity and nearly identical $z_{\rm{vtx}}$ position, refered to hereon as inclusive events.
\begin{figure}[htp]
\includegraphics[width=0.51\textwidth]{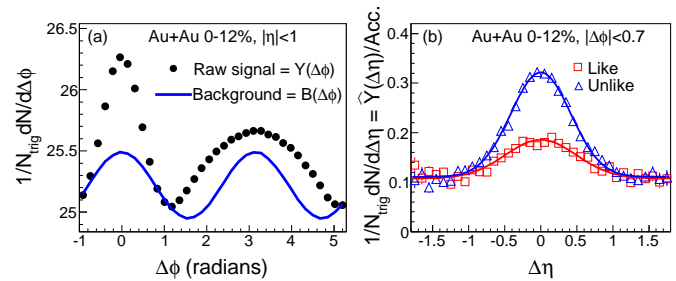}
\caption
{
Correlated hadron distribution in (a) $\Delta\phi$ ($\mid$$\eta$$\mid$$<$1),  
and (b) $\Delta\eta$ ($\mid$$\Delta\phi$$\mid$$<$0.7) with a high-$p_{\perp}$ trigger particle in 0-12$\%$ Au+Au collisions for 
3$<$$p_{\perp}^{\rm{(t)}}$$<$10 GeV/$c$ and 1$<$$p_{\perp}^{\rm{(a)}}$$<$3 GeV/$c$. 
The ZYA1-normalized flow background is shown in (a) by the curve.
The $\Delta\eta$ distributions in (b) are background subtracted and corrected for 
$\Delta\eta$ acceptance, and are for like- and unlike-sign
pairs separately. The curves in (b) are Gaussian fits. Errors are statistical.
}
\label{Fig1}
\end{figure}

Figure~\ref{Fig1}(a) shows the hadron $\Delta\phi$ distributions relative to the trigger particle in 0-12$\%$ 
Au+Au collisions. Also shown is the background 
$B$($\Delta\phi$) = $\it{a}F(\Delta\phi)\int^{\rm 2}_{\rm -2}B_{\rm{inc}}(\Delta\eta,\Delta\phi)d\Delta\eta$
where $B_{\rm{inc}}$ is constructed by mixing a trigger particle with associated 
particles from a different and inclusive event. 
The flow contribution 
\begin{equation}
{\it F}(\Delta\phi)=1+{\rm 2}v_{\rm 2}^{\rm (t)}v_{\rm 2}^{\rm (a)}\cos{({\rm 2}\Delta\phi)}+ {\rm 2}v_{\rm 4}^{\rm (t)}v_{\rm 4}^{\rm (a)}\cos{({\rm 4}\Delta\phi)}
\label{eq1}
\end{equation}
is added to mixed events using the measured, $\eta$-independent, $v_{2}$~\cite{flow} and a parameterization of 
$v_{4}$=1.15$v_{2}^{2}$~\cite{jason}.
A normalization factor, $a$, is applied to match the distribution in 
0.8$<$$\Delta\phi$$<$1.2, assuming zero yield at $\Delta\phi$$\sim$1 radian (ZYA1)~\cite{PRL}. 
The near-side ($\mid$$\Delta\phi$$\mid$$<$0.7) correlated hadron yield in 
$\Delta\eta$ is $\hat{Y}(\Delta\eta) = Y(\Delta\eta) - B(\Delta\eta)$,
where $Y(\Delta\eta)$ and 
$B(\Delta\eta)$ = $a\int^{\rm 0.7}_{\rm -0.7}B_{\rm{inc}}(\Delta\eta,\Delta\phi)F(\Delta\phi)d\Delta\phi$
are the signal and background distributions, respectively.
Figure~\ref{Fig1}(b) shows the $\hat{Y}(\Delta\eta)$ distribution, after 2-particle $\Delta\eta$ acceptance correction, for the 
like- and unlike-sign trigger-correlated particle pairs. 
Jet-like peaks at $\Delta\eta$$\sim$0 are observed, atop a broad, charge-independent 
pedestal (the ridge). A Gaussian fit to the peak yields $\sigma$=0.50$\pm$0.04 
for the like-sign and $\sigma$=0.41$\pm$0.01 for the unlike-sign pairs in $\Delta\eta$.\\
\indent All triplets of one trigger particle and two associated particles from the same event within $\mid$$\Delta\phi_{1,2}$$\mid$$<$0.7 are analyzed.
\begin{figure*}[htp]
\includegraphics[scale=0.82]{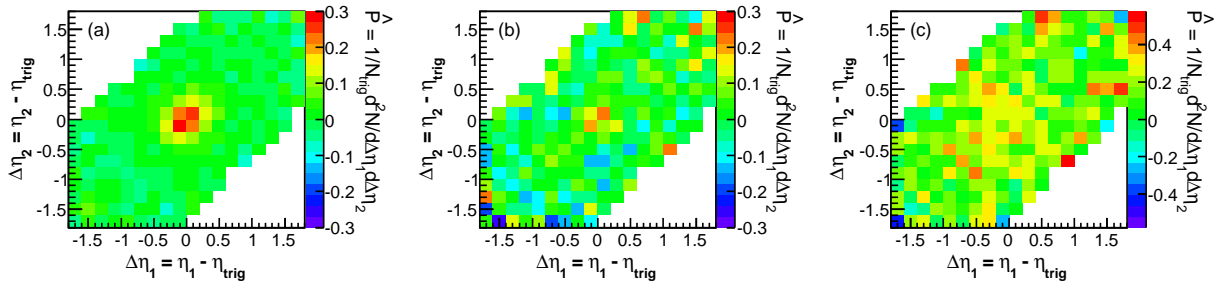}
\caption{
Background-subtracted charge-independent ($AAT$) correlated hadron pair density 
in (a) minimum bias $d$+Au, (b) 40-80$\%$ Au+Au, and (c) 0-12$\%$ Au+Au 
collisions for 3$<$$p_{\perp}^{\rm{(t)}}$$<$10 GeV/$c$ 
and 1$<$$p_{\perp}^{\rm{(a)}}$$<$3 GeV/$c$. The results are for near-side correlated hadrons within $\mid$$\Delta\phi_{1,2}$$\mid$$<$0.7, and corrected for the 3-particle 
$\Delta\eta$-$\Delta\eta$ acceptance. Statistical errors at ($\Delta\eta_{1}$,$\Delta\eta_{2}$)$\sim$(0,0) 
are approximately 0.033, 0.058, 0.084 for $d$+Au, 40-80$\%$ and 0-12$\%$ Au+Au, respectively.
}
\label{Fig2}
\end{figure*}
Combinatorial background $B_1$ (or $B_2$) arises where only one (or neither) 
of the two associated particles is correlated with the 
trigger particle besides flow correlation~\cite{back}.
The former cannot be readily obtained from the product of the event averaged 
$\hat{Y}(\Delta\eta)$ and $B(\Delta\eta)$, because of the varying $\Delta\eta$ acceptance from event 
to event. Instead, we construct $B_1$ by mixing trigger-associated pairs from the real 
event with a particle from a different and inclusive event, namely,
\begin{eqnarray}
B_{1} = \Bigl[a{Y(\Delta\eta_{\rm 1})B_{\rm inc}(\Delta\eta_{\rm 2})} \langle F^{\rm (t,2)}(\Delta\phi_{\rm 2}) + \notag \\ 
F^{\rm (1,2)}(\Delta\phi_{\rm 1} - \Delta\phi_{\rm 2}) + F^{\prime} - 1 \rangle \Bigr] + \Bigl[(1 \leftrightarrow 2) \Bigr] - \notag \\
\Bigl[2a^{2}{B_{\rm inc}(\Delta\eta_{\rm 1})B_{\rm inc}(\Delta\eta_{\rm 2})} \langle F^{\rm (t,1)}(\Delta\phi_{\rm 1}) + \notag \\ 
F^{\rm (t,2)}(\Delta\phi_{\rm 2}) + F^{\rm (1,2)}(\Delta\phi_{\rm 1} - \Delta\phi_{\rm 2}) + F^{\prime} - 2 \rangle \Bigr].
\label{eq2}
\end{eqnarray}
Here the last term is constructed by mixing the trigger particle with two different 
inclusive events to remove the uncorrelated part in the first two terms, and
\begin{eqnarray}
F^{\prime} & = & 2v_2^{\rm (t)}v_2^{(1)}v_4^{(2)}\cos(2\Delta\phi_{1}-4\Delta\phi_{2}) \notag \\
& + & 2v_2^{\rm (t)}v_2^{(2)}v_4^{(1)}\cos(4\Delta\phi_{1}-2\Delta\phi_{2})\notag\\
& + & 2v_2^{(1)}v_2^{(2)}v_4^{\rm (t)}\cos(2\Delta\phi_{1}+2\Delta\phi_{2}).
\label{eq3}
\end{eqnarray}
The flow terms~\cite{back} in $\langle \ldots \rangle$ are added in because they are lost in the event-mixing; 
their averages are taken within $\mid$$\Delta\phi_{1,2}$$\mid$$<$0.7.
The superscripts represent the $v_{2}$ and $v_{4}$ for trigger and
associated particles. To increase statistics, we mix each trigger particle  
with ten different inclusive events.\\
\indent The second background ($B_{2}$) is constructed by mixing a trigger particle with 
associated particle pairs from inclusive events thereby preserving 
all correlations between the two associated particles (denoted by $\otimes$)~\cite{back}: 
\begin{eqnarray}
B_{2} = a^2b\Bigl[B_{\rm inc}(\Delta\eta_{1})\otimes B_{\rm inc}(\Delta\eta_{2})\Bigr] \notag \\
\langle F^{\rm (t,1)}(\Delta\phi_{1})+F^{\rm (t,2)}(\Delta\phi_{\rm 2})+F^{\prime}-1\rangle.
\label{eq4}
\end{eqnarray}
The factor $a^{2}b$ scales the number of associated hadron pairs in the 
inclusive event to that in the background underlying the triggered event: 
$b = (\langle N(N-1)\rangle/\langle N \rangle^2)_{\rm bkgd}/(\langle N(N-1) \rangle/\langle N \rangle^2)_{\rm inc}$
where $N$ denotes the associated hadron multiplicity~\cite{back}.
If the associated hadron multiplicity distributions in both the inclusive
event and the background are Poissonian, or deviate from it equally, then $b$=1.
We obtain $b$ as follows. We scale the 
correlated hadron $\Delta\eta$ distribution such that there would be no ridge in 1.0$<$$\mid$$\Delta\eta$$\mid$$<$1.8, 
and this gives a new value for $a$. We repeat our analysis with this new $a$, and obtain
$b$ by requiring the average correlated hadron pair density  
in 1.0$<$$\mid$$\Delta\eta_{1,2}$$\mid$$<$1.8 be zero.
We use the obtained $b$ with the default ZYA1 $a$ to 
obtain the final 3-particle coincidence signal.
The assumption in this procedure is:
\begin{equation}
\Bigl[\langle N(N-1)\rangle/\langle N \rangle^2\Bigr]_{\rm bkgd} =\Bigl[\langle N(N-1)\rangle/\langle N \rangle^2\Bigr]_{\rm bkgd+ridge},
\label{eq5}
\end{equation}
and is reasonable gauged from multiplicity distributions of inclusive and triggered events.
The background-subtracted correlated pair density is corrected for 3-particle $\Delta\eta$-$\Delta\eta$ acceptance, which 
is obtained from event-mixing of a trigger particle with associated particles from two different 
inclusive events. We use ten pairs of inclusive events for each trigger particle in the mixing.\\
\indent The main sources of systematic uncertainty in our results 
are those in $a$, $b$ and $v_{2}$. 
These uncertainties are mostly correlated, therefore having insignificant effect 
on the shapes of our correlated density distributions.
The $a$ and $b$ values for 0-12$\%$ Au+Au collisions are 
$0.998^{+0.002}_{-0.001}$ (syst.) and $0.99986^{+0.00002}_{-0.00004}$ (syst.), respectively.
The uncertainty on $a$ is estimated by using the normalization
ranges of 0.9$<$$\Delta\phi$$<$1.1 and 0.7$<$$\Delta\phi$$<$1.3. 
That on $b$ is estimated by using the normalization 
ranges of 1.8$<$$\mid$$\Delta\eta$$\mid$$<$1.2 and 1.2$<$$\mid$$\Delta\eta$$\mid$$<$0.6. 
We note that the ridge is defined under the assumption of ZYA1 in $\Delta\phi$, 
by the factor $a$. Deviations of $a$ from this assumption are not included in our systematic
uncertainties. Such deviations ({\it e.g.} 3-particle ZYAM~\cite{jason}) do not introduce 
significant change to the shape of the ridge.\\ 
\indent The $v_{2}$ systematic range used in our analysis is given by those from 
the modified reaction plane and 4-particle cumulant methods~\cite{PRL} and
their average is used as our nominal $v_{2}$. An additional systematic uncertainty arises from possible correlation of the ridge with the reaction plane 
which is not included in Eq.~(\ref{eq2}). The estimated uncertainty from this source and that from $v_{2}$ are added in quadrature and referred 
to generally as flow uncertainty.\\
\indent Figure~\ref{Fig2}(a-c) shows the background subtracted charge independent (referred to as
$AAT$) correlated hadron pair density ($\hat{P}$) for minimum bias $d$+Au, 40-80$\%$ and 0-12$\%$ Au+Au collisions, respectively. 
The $d$+Au and 40-80$\%$ Au+Au results show a peak at ($\Delta\eta_{1},\Delta\eta_{2}$)$\sim$(0,0), consistent 
with jet fragmentation in vacuum. A similar peak is also observed in 0-12$\%$ Au+Au 
collisions, but it is atop an overall pedestal. This pedestal is composed of the ridge particle pairs, 
and does not seem to have other structures in ($\Delta\eta_{1},\Delta\eta_{2}$).
To see this quantitatively, Fig.~\ref{Fig3}(a) shows the average $\langle\hat{P}\rangle$ for $AAT$ as a function of $R$=$\sqrt{\Delta\eta_{1}^{2} + \Delta\eta_{2}^{2}}$.
The average density is peaked at $R$$\sim$0 and decreases with $R$ for all systems.
\begin{figure}[h]
\includegraphics[width=0.52\textwidth]{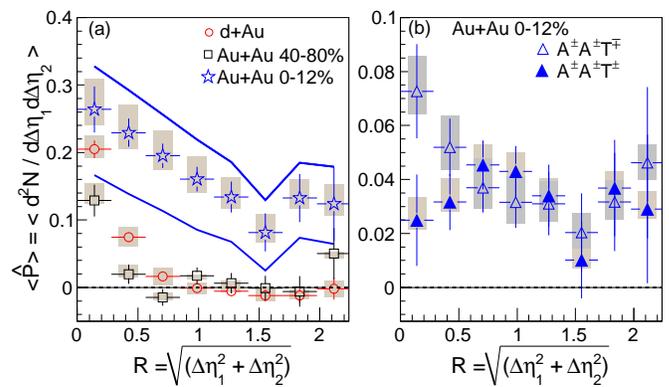}
\caption{
The average correlated hadron pair density per trigger particle as a function of $R$ (a) for all charges in minimum bias $d$+Au, 40-80$\%$ Au+Au 
and 0-12$\%$ Au+Au collisions, and (b) for same-sign associated particles 
($A^{\pm}A^{\pm}T^{\pm}$ and $A^{\pm}A^{\pm}T^{\mp}$) in 0-12$\%$ Au+Au collisions.
Systematic uncertainties are shown in the shaded boxes due to background 
normalization and in the solid curves due to flow.
}
\label{Fig3}
\end{figure}
For $d$+Au and 40-80$\%$ Au+Au collisions the average density at $R$$>$1 is consistent with zero, 
indicating no ridge contribution. On the other hand, in 0-12$\%$ Au+Au 
collisions, the average denstiy drops more slowly and becomes approximately constant above $R$$>$1, 
indicating the presence of the ridge.\\
\indent Jet fragmentation has a charge ordering property, as shown at $\mid$$\Delta\eta$$\mid$$\sim$0 
in Fig.~\ref{Fig1}(b). The probability to fragment into three 
same-sign hadrons at our energy scale is negligible~\cite{jet1,jet2}. Any 
correlation in three same-sign hadron triplets may therefore be interpreted 
as ridge correlation. Thus, we analyze our data with same-sign 
triplets ($A^{\pm}A^{\pm}T^{\pm}$) only, as well as with a same-sign 
associated pair and an opposite-sign trigger particle ($A^{\pm}A^{\pm}T^{\mp}$). 
The results are shown in Fig.~\ref{Fig3}(b). 
Indeed, no jet-like component is apparent in $A^{\pm}A^{\pm}T^{\pm}$. The $A^{\pm}A^{\pm}T^{\mp}$ 
result contains both jet-like and ridge components. The contribution from other charge combinations, 
namely $A^{\pm}A^{\mp}T^{\pm}$, are simply the difference between $AAT$ in 
Fig.~\ref{Fig3}(a) and ($A^{\pm}A^{\pm}T^{\pm}$ $+$ $A^{\pm}A^{\pm}T^{\mp}$) in Fig.~\ref{Fig3}(b). 
We found this to be equal to twice the $A^{\pm}A^{\pm}T^{\mp}$ contribution within errors.\\
\indent The ridge is very similar for like- and unlike-sign trigger-associated pairs
at $\mid$$\Delta\eta$$\mid>$0.7 as shown in Fig.~\ref{Fig1}(b), thus we expect the ridge contributions in the correlated pair density 
to be the same in all charge combinations. 
We verified this for large $\Delta\eta$ correlated pair densities within our current statistics, as can be
seen from Fig.~\ref{Fig3}(b).
Therefore the total ridge particle pair density ($\hat{P}_{rr}$) can be obtained as four times 
$A^{\pm}A^{\pm}T^{\pm}$.
The remaining jet-like signal, the sum of jet-like correlated particle pairs ($\hat{P}_{jj}$) and 
cross pairs of a jet-like and a ridge particle ($\hat{P}_{jr}$), can 
then be obtained by subtracting the total ridge from $AAT$.\\
\indent Figure~\ref{Fig4}(a) shows the $R$ dependence 
of the average $\langle\hat{P}_{rr}\rangle$ and $\langle\hat{P}_{jj}\rangle$+$\langle\hat{P}_{jr}\rangle$ in 0-12$\%$ Au+Au collisions.
The ridge pair density is consistent with a constant 0.14$\pm$0.02 ($\chi^{2}$/ndf=5.8/7).
Gaussian fits indicate a best fit value $\sigma$=2.1 ($\chi^{2}$/ndf=4.8/6, solid curve) and $\sigma$$>$1.4 
(dashed curve) with 84$\%$ confidence level.
On the other hand, the jet-like component is narrow with a Gaussian 
$\sigma$=$0.34^{+0.13}_{-0.09}$ ($\chi^{2}$/ndf=0.8/6, dominated by statistical errors), comparing well to those from the correlated single hadron density.\\
\begin{figure}[htp]
\includegraphics[width=0.52\textwidth]{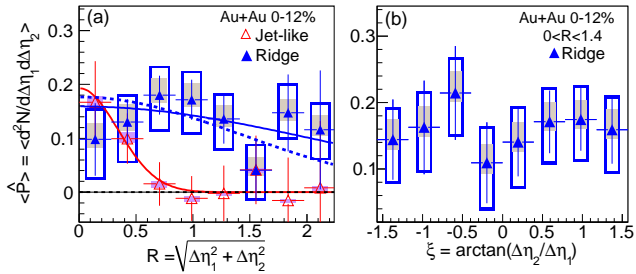}
\caption{
The average correlated hadron pair density per trigger particle in 0-12$\%$ Au+Au collisions 
(a) for the jet-like and ridge components as a function of $R$, and (b) for the 
ridge as a function of $\xi$ within $R$$<$1.4.
The solid curves are Gaussian fits. The dashed curve is a Gaussian fit
with a fixed $\sigma$=1.4 (see text) to the ridge data.
The systematic uncertainities on the ridge data are shown in shaded boxes due to background normalization and in open boxes due to flow.
}
\label{Fig4}
\end{figure}
\indent In order to investigate possible structures in the ridge, we show in
Fig.~\ref{Fig4}(b) the average ridge particle pair density
as a function of $\xi$=${\rm \arctan}(\Delta\eta_{2}/\Delta\eta_{1})$ within 
$R$$<$1.4. The data are consistent with a uniform 
distribution in $\xi$ ($\chi^{2}$/ndf=1.7/7). 
This suggests that the ridge particles are uncorrelated in $\Delta\eta$ 
not only with the trigger particle but also between themselves. 
In other words, the ridge appears to be uniform in $\Delta\eta$ event-by-event.\\
\indent Correlation between the jet-like correlated hadrons and the ridge would yield horizontal and vertical stripes in the correlated pair 
density in Fig.~\ref{Fig2}, resulting in a finite $\hat{P}_{jr}$, a non-zero signal in Fig~\ref{Fig4}(a) at large $R$. We found 
$\langle\hat{P}_{jr}\rangle$=$-0.001\pm0.030$, averaged over the $\mid$$\Delta\eta_{1}$$\mid$$<$0.7 and $\mid$$\Delta\eta_{2}$$\mid$$>$0.7 
region and its mirror region. On the other hand for the correlated jet-jet and ridge-ridge pairs, 
$\sqrt{\langle\hat{P}_{jj}\rangle\langle\hat{P}_{rr}\rangle}$=$\sqrt{(0.081\pm0.034)\times(0.114\pm0.039)}$=$0.096\pm0.026$ where the averages 
are taken with $\mid$$\Delta\eta_{1,2}$$\mid$$<$0.7 and $\mid$$\Delta\eta_{1,2}$$\mid$$>$0.7, respectively. The comparision between these 
two pair density magnitudes (whose systematic uncertainities are strongly correlated) suggests that production of the ridge and production 
of the jet-like particles may be uncorrelated.\\
\indent Our data qualitatively distinguish between some of the ridge models.
(i) Longitudinal flow~\cite{nestor} would push correlated particles
in one direction yielding a diagonal excess in $\Delta\eta$-$\Delta\eta$, 
disfavored by the present data. 
(ii) Turbulent color fields~\cite{majumdar} would generate 
a broad ridge in $\Delta\eta$, which may however still be too narrow to reconcile with the width of our ridge pair density distribution.
(iii) Recombination between thermal and shower partons~\cite{hwa} 
should produce horizontal and vertical stripes in correlated pair density distribution 
which is disfavored by the data, and it does not have a mechanism for long range
$\Delta\eta$ correlations.
(iv) The momentum kick model incorporates a broad ridge
as input, but it should produce a much larger ridge on the 
away-side than on the near-side which is not supported by data~\cite{marco}, and also
may not describe other data such as
the reaction plane dependence of the ridge 
in di-hadron correlations~\cite{aoqi}.
(v) QCD bremsstrahlung~\cite{voloshin,shuryak} or 
color flux tube fluctuations~\cite{dumitru,dusling,jun} 
would yield a structure-less pair density~\cite{dusling} for the ridge as observed in our data, however the 
correlations between jet-like particles and ridge, as 
expected from these models, are not observed with our present sensitivity.
Clearly more quantitative model calculations are needed to compare to the data 
reported here and elsewhere~\cite{PRL,joern,aoqi} to further our understanding 
of the ridge.\\
\indent In summary, we have presented the first 3-particle coincidence measurement in $\Delta\eta$-$\Delta\eta$ in minimum bias $d$+Au, 40-80$\%$ 
and 0-12$\%$ Au+Au collisions at $\sqrt{{\it s}_{NN}}$ = 200 GeV. 
The $p_{\perp}$ ranges are 3$<$$p_{\perp}^{\rm{(t)}}$$<$10 GeV/$c$ for the trigger particle 
and 1$<$$p_{\perp}^{\rm{(a)}}$$<$3 GeV/$c$ for both associated particles.
A correlated hadron pair density peak at ($\Delta\eta_{1}$,$\Delta\eta_{2}$)$\sim$(0,0), 
characteristic of jet fragmentation, is observed in all systems.
This peak sits atop a broad pedestal in 0-12$\%$ Au+Au collisions, 
which is composed of particle pairs from the ridge.
We have exploited the charge ordering properties to separate the jet-like and  
ridge components. 
We found that same-sign associated pairs correlated with a same-sign 
trigger particle are dominated by the ridge.
While the jet-like particle pair density is narrowly confined, the ridge is broadly distributed and is approximately 
uniform in $\Delta\eta$. A Gaussian fit in $R$ to the average correlated pair density of the ridge yields $\sigma$$>$1.4 with 84$\%$ confidence level.
Except for the correlations at $\Delta\phi$$\sim$0, the particles from the ridge appear to 
be uncorrelated in $\Delta\eta$ not only with the trigger particle, but also between themselves; 
they are uniform in our measured $\Delta\eta$ range event-by-event. 
No correlation is found between production of the ridge and production of the jet-like particles, suggesting the ridge may be formed from the bulk medium itself.\\
\indent We thank the RHIC Operations Group and RCF at BNL, the NERSC Center at LBNL and 
the Open Science Grid consortium for providing resources and support. This work was 
supported in part by the Offices of NP and HEP within the U.S. DOE Office of Science, 
the U.S. NSF, the Sloan Foundation, the DFG cluster of excellence 
`Origin and Structure of the Universe', CNRS/IN2P3, STFC and EPSRC of the United Kingdom, 
FAPESP CNPq of Brazil, Ministry of Ed. and Sci. of the Russian Federation, NNSFC, CAS, 
MoST, and MoE of China, GA and MSMT of the Czech Republic, FOM and NWO of the 
Netherlands, DAE, DST, and CSIR of India, Polish Ministry of Sci. and Higher Ed., 
Korea Research Foundation, Ministry of Sci., Ed. and Sports of the Rep. Of Croatia, 
Russian Ministry of Sci. and Tech, and RosAtom of Russia.

\end{document}